# Electronic and phonon excitations in α-RuCl₃

S. Reschke,[1] F. Mayr,[1] Zhe Wang,[1,2] Seung-Hwan Do,[3] K.-Y. Choi,[3] and A. Loidl[1,*]

[1]Experimental Physics V, Center for Electronic Correlations and Magnetism, University of Augsburg,
86135 Augsburg, Germany
[2]Institute of Radiation Physics, Helmholtz-Zentrum Dresden-Rossendorf,
01328 Dresden, Germany
[3]Department of Physics, Chung-Ang University, Seoul 06974, Republic of Korea

We report on THz, infrared reflectivity and transmission experiments for wave numbers from 10 to 8000 cm⁻¹ (~ 1 meV – 1 eV) and for temperatures from 5 to 295 K on the Kitaev candidate material α-RuCl₃. As reported earlier, the compound under investigation passes through a first-order structural phase transition, from a monoclinic high-temperature to a rhombohedral low-temperature phase. The phase transition shows an extreme and unusual hysteretic behavior, which extends from 60 to 166 K. In passing this phase transition, in the complete frequency range investigated we found a significant reflectance change, which amounts almost a factor of two. We provide a broadband spectrum of dielectric constant, dielectric loss and optical conductivity from the THz to the mid infrared regime and study in detail the phonon response and the low-lying electronic density of states. We provide evidence for the onset of an optical energy gap, which is of order 200 meV, in good agreement with the gap derived from measurements of the DC electrical resistivity. Remarkably, the onset of the gap exhibits a strong blue shift on increasing temperatures.

## I. INTRODUCTION

Quantum spin liquids (QSLs) are a fascinating playground in modern solid-state physics with unusual magnetic properties and exotic spin excitations [1]. In QSLs electron spins show strong and long-range entanglement, but do not undergo magnetic ordering, even in the 0 K limit. One decade ago, Kitaev [2] proposed an exactly solvable model for S =1/2 Ising spins on a two-dimensional (2D) honeycomb lattice, with bond-directional anisotropic interactions, which result in a topological QSL and emergent Majorana fermions. In a seminal work, Jackeli and Khaliullin proposed to realize Kitaev-type interactions in materials with strong spin-orbit coupling [3]. Despite enormous experimental efforts, so far a pure Kitaev-type QSL, could not be identified experimentally, mainly due to the fact that further, even marginal magnetic interactions drive the spin system on the 2D lattice into long-range antiferromagnetic (AFM) order [4]. Prime candidates of Mott insulators with honeycomb-based structures were some layered iridates [5,6]. Quite recently, α-RuCl₃, which forms an almost ideal two-dimensional honeycomb lattice with very weak interlayer coupling, was proposed to be a prime candidate for the realization of Kitaev physics. It is known since half a century that α-RuCl₃ is an antiferromagnet at low temperatures [7,8] and hence exhibits three-dimenional (3D) magnetic order, preempting the exotic QSL ground state. However, fractionalized quasiparticles, composed of itinerant and localized Majorana fermions [9,10,11], still could exist in a temperature regime above magnetic order. Indeed, recent Raman [12,13,14] and neutron scattering experiments [15,16,17] provided experimental evidence of fractionalized excitations at low temperatures reminiscent of Kitaev physics. Another possible explanation was put forth recently by Winter et al. [18]: The experimentally observed broad continua could be described by incoherent magnetic excitations originating from strong magnetic anharmonicity, due to the presence of spin-orbit coupling and frustrated anisotropic magnetic interactions.

After early reports on synthesis and structure of α-RuCl$_3$ [19], Fletcher et al. [7] reported synthesis, structural and magnetic characterization of α- and ß-RuCl$_3$. This sample characterization was detailed in Ref. [8] and complemented by room-temperature infrared (IR) experiments. The structural units of α-RuCl$_3$ are honeycomb layers of ruthenium, separated by two hexagonal layers of chlorine. Ru$^{3+}$ (4d$^5$) is coordinated by Cl$^-$ ions in almost ideal octahedral symmetry with a slight monoclinic distortion. The crystal-field (CF) splitting is large, leading to a t$_{2g}$ ground state with five electrons in low spin configuration. The ruthenium layers, sandwiched between two hexagonal layers of chlorine, represent strongly bonded stacks and are connected to neighboring layers by weak Van der Waals (VdW) forces only. Due to the very weak VdW forces between the molecular of these infinite layer compounds, stacking faults quite naturally appear. Hence, some controversy exists in literature about the structure of α-RuCl$_3$. After early structural studies, which resulted in a highly symmetric P3$_1$12 space group with nearly isotropic edge-sharing RuCl$_6$ octahedra [19,7,8], it seems now well established that the room temperature symmetry is monoclinic with space group C2/m [20,21,22], which is the same as the structure of the related infinite-layer compound CrCl$_3$ at 300 K [23]. Most of the recent publications on α-RuCl$_3$ assume that the low-temperature structure remains the same monoclinic C2/m structure. However, many related infinite layer compounds undergo a further transition into a low-temperature rhombohedral structure [23,24,25]. Of course, concerning the Kitaev spin-liquid state, the detailed knowledge of the structural ground state in α-RuCl$_3$, possibly with the ideal stacking sequence, is of prime importance. Three recent publications provide detailed reports on a structural phase transition in RuCl$_3$, with a strongly hysteretic behavior located around 150 K [14,26,27]. Structural details and a sequence of critical temperatures were reported in [27]. The phase transition was also mapped out via Raman spectroscopy [14]. A similar phase transition, induced by changes of the molecular stacking sequence, also appears in chromium tri-chloride at 240 K [23, 25], in chromium tri-iodide close to 220 K [24], and indicates a transition into a rhombohedral structure with $R\bar{3}$ symmetry. In the low-temperature structure, the chlorine ions are almost hexagonal close-packed (hcp), with an AB-type stacking sequence, in contrast to the nearly face-centered cubic (fcc) close-packing arrangement in the high-temperature monoclinic form, representing an ABC stacking sequence. This rhombohedral-to-monoclinic transition can occur by a simple translation of neighboring sandwiches along one direction [23] and is closely related to the onset of magnetic order at low temperatures as will be discussed later.

In α-RuCl$_3$, there exists a close correlation between structural details and low-temperature AFM order. Magnetic ordering temperatures between 6.5 and 15.6 K were reported [8,27,28,29,30,31]. Sometimes even a sequence of magnetic phase transitions occurred between 7 and 17 K [26]. It seems that the stacking sequence is closely linked to the onset of magnetic order. AB type crystals show a sharp transition close to 7 K, while crystals with an ABC-type stacking order undergo a broad magnetic transition close to 14 K [31]. Hence, crystals undergoing this monoclinic to rhombohedral phase transition will exhibit an almost ideal stacking sequence and the lowest magnetic transition temperatures. The interpretation of neutron diffraction experiments tends to converge in a zigzag spin structure, with the spins tilted from the ab plane and with AFM stacking along the crystallographic c direction [27,30,31].

Fletcher et al. [8] reported the first optical data on α-RuCl$_3$. At room temperature infrared (IR) bands close to 188, 274, and 315 cm$^{-1}$, corresponding to 23.3, 34.0 and 39.1 meV, were identified. Later on, Binotto et al. [32] and Guizzetti et al. [33] reported further optical investigations on α-RuCl$_3$. Because of the strongly two-dimensional character of the molecular stacks, D$_{3d}$ symmetry was assumed for the isolated sandwich of the layered structure. Group factor analysis then gives 5 IR active modes, namely 2 A$_{2u}$ modes, polarized parallel to the c axis, which is the stacking direction and 3 E$_u$ modes polarized perpendicular to the c axis, within the stacks. In this early work, in the far-infrared (FIR) reflectivity spectra one optical mode with a transverse eigenfrequency close 300 cm$^{-1}$ (37.2 meV) was observed [33]. Utilizing transmission experiments, Refs [32] and [33] also reported IR forbidden d-d transitions at 290, 510 and 710 meV within the ruthenium 4d bands. A detailed study of the temperature dependence of the polarized Raman response is presented in Ref. [12]. In addition to a broad scattering continuum at low energies, these authors observed a number of narrow phonon lines between 10 and 40 meV, which were interpreted in terms of an isolated layer of D$_{3d}$ symmetry. In the temperature dependence, these authors found no indications of a structural phase transition [12]. In contrast, a detailed analysis of Raman shifts, widths and intensities indicated a structural phase transition with broad hysteresis [14]. In addition, these authors interpreted the broad continuum at low frequencies as Kitaev magnetism concomitant with fractionalized excitations [14].

α-RuCl$_3$ is a semiconducting compound and measurements of the DC conductivity provide some insight on the size of the band gap close to the Fermi energy. Binotto et al. [32] determined the electrical conductivity and observed conventional semiconducting behavior, which was found to be four orders of magnitude larger when measured perpendicular to the planes, than within the planes. However, in both cases and below room temperature they observed a band gap of the order of 100 meV. Later on Rojas and Spinolo [34] measured DC conductivity and Hall coefficient and found a band gap between valence and conduction band of order 300 meV. These authors argued that this band gap is very close to the first absorption band, detected in the transmission experiments mentioned above and hence, that this lowest on-site atomic-like d-d excitation corresponds to a transition from the valence to the conduction band.

Later on, Sandilands et al. [35,36] investigated the electronic structure of α-RuCl$_3$ in detail. In good agreement with earlier results [32,33], these authors identified orbital excitations from the t$_{2g}$$^5$ ground state to an intermediate spin state t$_{2g}$$^4$e$_g$ close to 0.3, 0.5 and 0.7 eV. Above 1 eV the optical response is dominated by intersite ruthenium d-d excitations followed by charge-transfer excitations. Utilizing Raman spectroscopy, the authors of Ref. [35] also identified the transitions from the J = 3/2 ground quartet into the J = 1/2 excited doublet, within the spin-orbit split t$_{2g}$ ground state. The splitting amounts 3λ/2, with λ being the spin-orbit coupling (SOC) constant. From these results, the authors were able to deduce the value of the spin-orbit coupling approximately as λ ∼ 100 meV. Plumb et al. [37] studied the electronic structure of α-RuCl$_3$ by combined x-ray absorption and optical spectroscopy and, over all, identified a similar electronic density of states, with the Ru t$_{2g}$ − e$_g$ transitions ranging from 1 to 3 eV, followed by charge-transfer transitions from chlorine p into ruthenium e$_g$ states. The main difference of these interpretation in terms of band structures [37], compared to the single ion interpretation presented in Ref. [33, 35,36], concerns the excitations below 1 eV: Only the inclusion of moderate electronic correlations can establish an insulating ground state and the dominating structure in the dielectric loss function then signals the transition across the Mott-Hubbard gap. Taking into account electronic correlations with a Hubbard U ~ 1.5 eV induces a gap in the density of states at the Fermi level and splits the density of states of the $t_{2g}$ level into a broad lower and a narrow upper Hubbard band, with a transition energy of about 250 meV [37], in good agreement with the conductivity results [34].

Based on recent band-structure calculations [30,37,38], one can conclude that the crystal-field splitting of the ruthenium d bands is of order 2 eV, with the Fermi energy still within the $t_{2g}$ band, even when assuming spin-orbit coupling. Energy separation of the upper band edge of the $t_{2g}$ states and the lower band edge of $e_g$, is of order 1 eV. Electronic correlations will induce a Mott-Hubbard gap within the $t_{2g}$ states, which is expected to be of order 300 meV (~ 2600 cm$^{-1}$) [37], well below the crystal-field splitting. These electronic transitions between the lower and upper Hubbard band, obviously are almost invisible in IR reflectivity, but are certainly detectable in transmission experiments. It is unclear, whether in addition to this transition across the correlation gap, single-ion excitations between ruthenium d levels, as reported in literature [32,33,35], still can be identified experimentally. To complete this survey on optical experiments on α-RuCl$_3$, very recently, the THz conductivity has been reported by Little et al. [39] for frequencies between 1 and 7 meV. These authors observed an absorption continuum above a sharp threshold energy of 1 meV. In addition, they observed a well-defined magnetic dipole transition at 2.5 meV, in the magnetically ordered phase. THz experiments as function of temperature and magnetic field were also performed by Wang et al. [40]. Below 60 K these authors observed well defined continua, which could correspond to fractionalized excitations in addition to a well-defined magnetic excitation. Close to the quantum phase transition, which appears at a critical magnetic field of 7 T, this magnetic continuum dominates the dynamic response.

In this optical investigation of α-RuCl$_3$ we provide detailed IR reflectivity and transmission experiments ranging from 10 cm$^{-1}$ up to 8000 cm$^{-1}$, corresponding to approximately 1.2 meV – 1.0 eV in energy and for temperatures from 5 K up to room temperature. We were interested in the temperature evolution of phonon eigenfrequencies and intensities, to identify possible structural phase transitions, and aiming at a detailed study of electronic density of states (DOS) well below the crystal field-split energy of ~ 1 eV. The latter, to identify the nature of the low-energy electronic excitations, as being either of single-ion or of band type. In addition, we wanted to present a broadband spectrum of dielectric constant and dielectric loss in α-RuCl$_3$, covering frequencies from the THz up to the mid IR (MIR) regime, to gain some information on low-lying excitations, phonon modes, and on the low-energy density of electronic states and their possible interplay. In addition, the optical response of QSLs seems to be of reasonable scientific interest; e.g., a spin-induced optical conductivity with a power-law dependence in the THz regime was reported in the QSL candidate Herbertsmithite [41]. However, a detailed study of the temperature and field dependence of excitations in the THz regime of α-RuCl$_3$ is not in the focus of this work and will be published separately [40].

## II. EXPERIMENTAL DETAILS

High-quality α-RuCl$_3$ single crystals, which were used in the various optical experiments of this work, were grown by a vacuum sublimation method. Details of sample preparation and characterization are described in Ref. [17]. It is interesting to note that the magnetic ordering temperature of 6.5 K belongs to the lowest ever reported for α-RuCl$_3$, indicating an ideal AB stacking sequence. The time-domain THz transmission experiments using a TPS Spectra 3000 spectrometer (TeraView Ltd.) were performed with the THz wave vector perpendicular to the crystallographic *ab* plane for frequencies from approximately 1.5 – 12 meV in a He-flow cryostat (Oxford Optistat) between liquid-helium and room temperature. Time-domain signals were obtained for reference (empty aperture) and samples from which the power spectra were evaluated via Fourier transformation. The sample was a platelet of size $3 \times 3$ mm$^2$, with a thickness of 0.9 mm. Different samples of thicknesses ranging from 0.1 to 0.4 mm were used to measure the IR transmission for energies ranging from 100 to 8000 cm$^{-1}$, using the Bruker FT-spectrometers IFS113v and IFS66v/S with appropriate sets of sources, beam splitters and detectors. He-flow cryostats from CryoVac were utilized to vary the sample temperature from 5 to 295 K. The same combination of spectrometers and cryostats was also employed to acquire reflectivity spectra of the thicker sample (d = 0.9 mm) in the same frequency and temperature range. In all IR experiments the incident wave was perpendicular to the molecular stacks, parallel to the crystallographic $c$ direction. In this optical reflectivity setup, a gold mirror was used as a reference. The reflectivity spectra were converted into the complex dielectric permittivity $\varepsilon(\omega)$ with the Kramers–Kronig constrained variational method, developed by Kuzmenko [42], which is included in the RefFIT program [43]. With this approach, it is possible to obtain $\varepsilon(\omega)$ without the need of specific extrapolations at the low- and high-frequency edges of the measured spectra.

## III. RESULTS AND DISCUSSION

### A. Reflectivity changes at the phase transition

Figure 1 shows the reflectivity measured on $\alpha$-RuCl$_3$ for wave numbers between 100 and 600 cm$^{-1}$ and temperatures between 5 and 300 K, with the incident light perpendicular to the molecular stacks. These measurements were performed on the single crystalline sample with thickness of about 0.9 mm and the data shown are measured on cooling. At all temperatures, the reflectivity is characterized by one strong absorption band just above 300 cm$^{-1}$. As function of temperature, this reststrahlen band neither undergoes significant shifts, nor does it show any indications of mode splitting. However, it reveals a tremendous and highly unusual temperature dependence of the absolute values of the reflectivity. On decreasing temperatures, the maximum reflectivity in the middle of this reststrahlen band increases down to 120 K, exhibits a step-like drop down to 60 K, and then remains constant on further decreasing temperatures. A further unusual behavior concerns the wave number dependence as function of temperature. At room temperature, the spectrum looks rather conventional as in a variety of 3D solids with one IR active optic mode in the FIR range: the reststrahlen band exhibits a smooth increase and a well-defined ionic plasma frequency, defining transverse and longitudinal optical phonon frequencies, respectively. At room temperature, a fit using a standard Lorentz oscillator, yields a transverse eigenfrequency of 311.5 cm$^{-1}$, a longitudinal eigenfrequency of 340.2 cm$^{-1}$ and a relatively low damping constant of 4.5 cm$^{-1}$. On decreasing temperatures, the reflectivity for the complete frequency range shows a strong temperature dependence and a number of wiggles develop, which are only weakly temperature dependent. The origin of this structure remains unclear. However, in different runs, on cooling as well as on heating, these modulated reflectivity structures are reproducible. It could be due to interference phenomena of molecular layers or of stacking faults.

We extended these reflectivity experiments to 7000 cm$^{-1}$. Representative results are shown in the inset of Fig. 1. In this wave number regime, the reflectivity is almost featureless, with no detectable electronic transition within experimental uncertainty. Based on estimates of the band gap from DC resistivity experiments [32,34], in this wave number, respectively energy regime, well below the onset of CF transitions, we expect signatures of electronic transitions from the valence to the conduction band, which obviously are below experimental uncertainty. It is highly unusual to observe no indications of the dominating optical band gap in reflectivity experiments. However, we observe a dramatic over-all temperature dependence of the reflectivity in this frequency regime, similar to the effects observed at phonon frequencies.

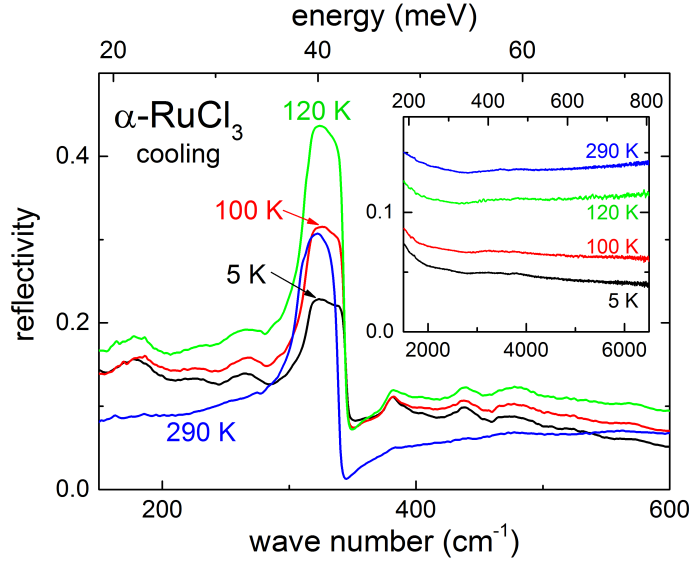

FIG. 1: Reflectivity of α-RuCl$_3$, measured with incident light perpendicular to the molecular stacks, between 100 and 600 cm$^{-1}$ (20 − 70 meV), for selected temperatures between 5 and 290 K. The measurements were performed on cooling. The inset shows the reflectivity between 1000 and 7000 cm$^{-1}$, where low-lying electronic excitations are expected, which is markedly featureless, but exhibits a strong and unusual overall temperature dependence.

In a first step, we wanted to unravel this strong temperature dependence, which is visible at all wavenumbers up to almost 1 eV. As representative examples, Fig. 2 shows the temperature dependence of the reflectivity as measured at 325 cm$^{-1}$, just in the middle of the reststrahlen band and at 4000 cm$^{-1}$, in the MIR regime, with its almost constant and featureless reflectivity (see inset in Fig. 1). In both cases, on heating and cooling, we observe a strongly hysteretic temperature dependence, indicative of a first order phase transition. On cooling, the reflectivity drops significantly at a temperature T* ∼ 120 K and tails off towards lower temperatures. On heating, at a temperature T$_{s2}$ ∼ 60 K, the hysteresis between cooling and heating opens, and at T$_{s1}$ ∼ 164 K the hysteresis closes and the reflectivity recovers its high-temperature value. These three characteristic temperatures are reproduced at very different wave numbers, as impressively documented in Figs 2 (a) and 2(b) for measurements at 325 and 4000 cm$^{-1}$. These hysteresis loops are perfectly reproducible and were measured at different wave numbers and on different single crystals. In the complete frequency range investigated, the reflectivity is drastically enhanced by more than a factor of two when passing from the low-temperature rhombohedral to the high-temperature monoclinic phase. This effect must mainly result from the stacking sequence. Similar colossal effects in reflectance are observed in phase-change materials, where the structure changes from crystalline to amorphous [44] and where the difference in reflectance is attributed to a significant change in bonding [45]. It could be that in the case of α-RuCl$_3$ stacking order and stacking sequence play a fundamental role in determining the overall reflectance.

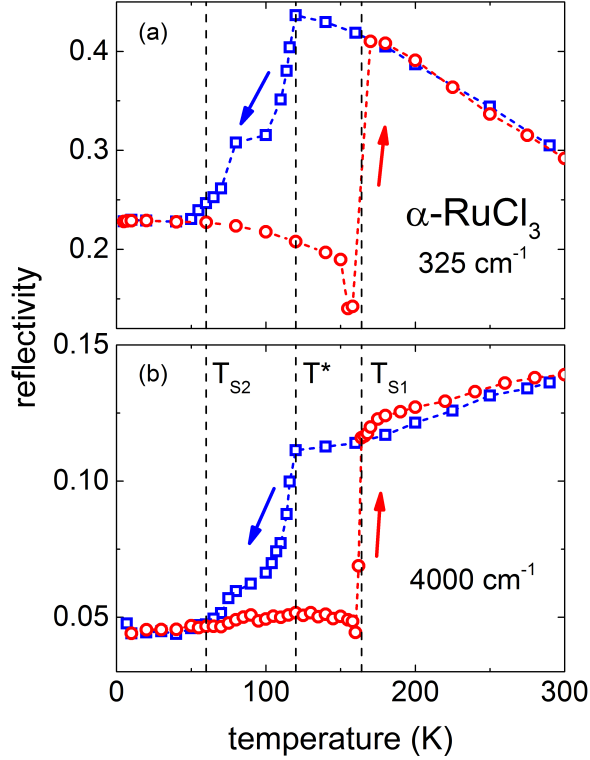

FIG. 2: Temperature dependence of the reflectivity in α-RuCl$_3$ with incident light perpendicular to the molecular stacks measured at (a) 325 cm$^{-1}$, in the middle of the reststrahlen band and (b) at 4000 cm$^{-1}$ in the MIR regime. Data are shown for heating (circles) and cooling (squares). The arrows indicate heating and cooling directions. The dashed vertical lines indicate characteristic temperatures as discussed in the text.

Prior to this work, a similar sequence of characteristic temperatures has been identified in the temperature dependence of lattice parameters measured by single-crystal x ray diffraction [27]. In these experiments, $T_{s1} \sim 166$ K and $T_{s2} \sim 60$ K, correspond to the characteristic temperatures of the first-order structural phase transitions with pronounced thermal hysteresis, indicated by length changes of the c axis, when the sample transforms from the monoclinic high-temperature to the rhombohedral low-temperature phase. On cooling, it seems that the monoclinic phase – viewed via the temperature dependence of the c axis - can be supercooled down to a temperature T*~ 115 K and then the c axis gradually approaches the low-temperature rhombohedral value. However, the in-plane lattice parameters show thermal hysteresis only between the characteristic temperatures $T_{s1}$ and T* [27]. Hence, the dramatic changes in reflectivity in α-RuCl$_3$ as documented in Fig. 2, results from changes in the c axis, probably induced by changes in the stacking sequence. On cooling, the stacking sequence rearranges continuously between T* and $T_{s2}$, while on heating this happens abruptly at $T_{s1}$. Figure 2 should be used as a recipe for the search for fractionalized excitations at low-temperatures: Any analysis should focus on temperatures below $T_{s2} \sim 60$ K to avoid any influences of purely structural effects.

**B. Phonon properties**

Assuming D$_{3d}$ symmetry of an isolated Cl-Ru-Cl sandwich only, we expect five IR allowed transitions, namely 2 $A_{2u}$ and 3 $E_u$ modes, with the electric field parallel and perpendicular to the c axis, respectively. Assuming the crystallographic monoclinic structure of the high-temperature phase 4 $A_u$ modes within the plane and 5 B$_u$ modes perpendicular to the plane are to be expected [46]. In the configuration used in the present experiments, we expect to observe at least the 3 E$_u$ modes, however, in reflectivity only 1 mode can be detected (see Fig. 1). To gain further insight, we performed transmission experiments on a single crystal with a thickness of 0.4 mm. The results are presented in Fig.3, which documents the transmission as measured in α-RuCl$_3$ between 100 and 2000 cm$^{-1}$ at a series of temperatures between 10 and 295 K. First, we discuss the phonon response below 1000 cm$^{-1}$. At all temperatures, we clearly observe three modes with relatively significant absorption, located close to 180, 300 and 600 cm$^{-1}$. There is some experimental evidence for a further phonon-like excitation close to 120 cm$^{-1}$ with weak absorptivity and strong damping. A clear fingerprint of this mode was also detected by our THz experiments (see later). The observation of 4 in-plane modes would not be compatible with the isolated sandwich picture of the molecular stacks, but rather would imply the importance of the monoclinic overall symmetry. However, there is a further problem: In all vibrational spectra reported for this tri-chloride structure, no eigenfrequencies beyond 400 cm$^{-1}$ were reported. This will be discussed later in more detail. Already at first sight, it is clear that the transmission is strongly temperature dependent. This is specifically true for energies beyond 80 meV and documents that multiphonon scattering will play an important role at FIR frequencies, but also that electron-phonon coupling must be extremely strong, responsible for a significant shift of the band edge. The extremely strong temperature dependence of multiphonon infrared absorption has been elucidated in seminal works long time ago in full detail [47,48,49].

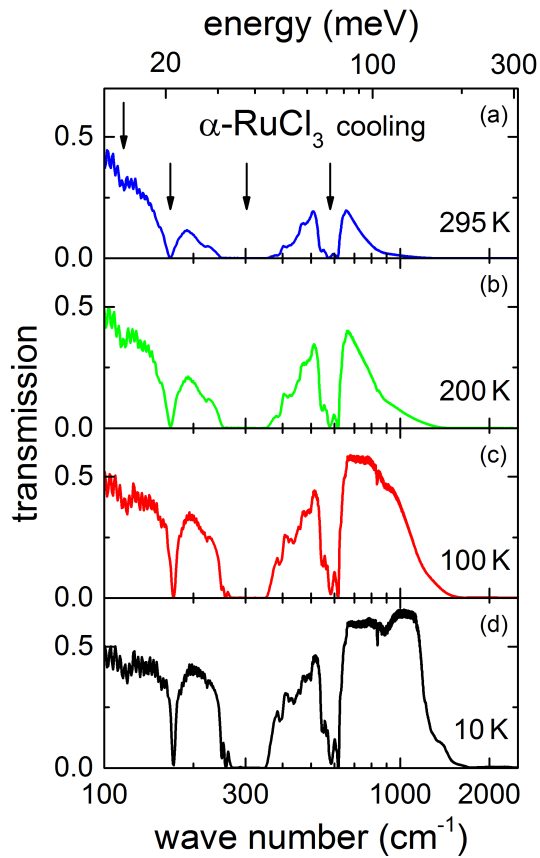

FIG. 3: IR transmission as measured in α-RuCl$_3$ for wavenumbers between 100 and 2000 cm$^{-1}$ and with the incident light perpendicular to the molecular stacks for a series of temperatures: (a) 295 K, (b) 200 K, (c) 100 (K), and (d) 10 K. All spectra shown have been measured on cooling. The transmittance is shown on a semilogarithmic plot. Arrows in (a) indicate eigenfrequencies of the phonon absorption bands.

The fine structure in the transmission, which is visible at all temperatures, results from multi-phonon scattering processes. Of course, this multi-phonon scattering could hide further weak one-phonon absorption processes. The overall transmission is strongly reduced on increasing temperatures, again due to the increase of allowed multi-phonon processes. All the observed modes are characterized by eigenfrequencies with a seemingly weak temperature dependence. There is no evidence of the appearance of extra modes or of mode splittings, indicative of any symmetry changes at a structural phase transition. This observation makes clear that the hysteretic phase transition observed in the temperature dependence of the over-all reflectivity mainly concerns the layering of molecular stacks or is of electronic origin only. On the other hand, indeed, the phonon properties are solely determined by the local $D_{3d}$ symmetry of each stack, while it seems that the overall symmetry, plays a minor role only.

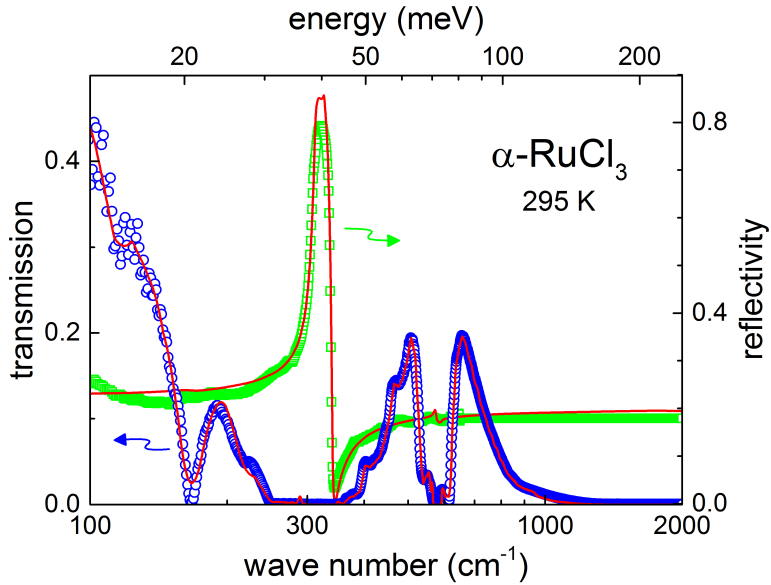

FIG. 4: Experimentally observed transmission (open circles) and reflectivity (open squares) vs. wavenumber in α-RuCl₃ at room temperature. These experiments have been performed with the incident light perpendicular to the molecular stacks. The solid lines corresponds to fits with a Kramers-Kronig constrained variational method following Ref. [42]. For these fits, the reflectivity data had to be scaled by a factor of 2.58.

As consistency check of our results on α-RuCl₃, which were obtained in transmission and reflection scattering geometry, we show in Fig. 4 transmission and reflectivity results as observed between 100 and 2000 cm⁻¹ at room temperature. Both data sets have been fitted with a Kramers-Kronig constrained variational method [42]. The results are shown as solid lines in Fig. 4. We achieve even quantitative very good agreement between the results obtained in transmission and reflection, when the reflectivity data are scaled by a factor 2.58. It is clear that the reflectivity results suffer from the imperfect surface of the as-grown single crystals. The very weak VdW-type bonding of the molecular stacks and the concomitant softness of the crystalline samples, do not allow any polishing procedure to arrive at surfaces with optical quality. The absolute values of the original as-measured reflectivity data are documented in Fig. 1. The modelling documents that indeed, in reflectivity only the phonon mode close to 300 cm⁻¹ can be detected. All the other phonon modes, as well as electronic transitions beyond 1000 cm$^{-1}$, have too low electric dipolar weight to be identified within the given experimental uncertainties. We think that the number of satellite peaks, which occur in the transmission are due to multi-phonon processes. At present, we do not provide any detailed analysis of possible multi-phonon modes.

To arrive at a quantitative estimate of the phonon properties, we analyzed the temperature dependence of the absorptivity assuming a simple Lorentzian-type oscillator model for all phonons, with eigenfrequency $\omega_0$, damping $\gamma$, and oscillator strength $\Delta\varepsilon$. We have to admit that while the quality of the fits is rather good at enhanced temperatures, in the high-temperature monoclinic phase, it gets increasingly worse on lowering the temperatures. For room temperature and for 10 K the results of these fits are listed in Tab. 1. Here the eigenfrequency corresponds to the frequency of the transverse optical modes. Only the mode close to 310 cm$^{-1}$ has considerable dipolar weight and hence, is detectable in IR reflectance measurements. The two high-frequency modes exhibit low damping, while the two low-frequency modes exhibit strong damping coefficients, with the damping of the lowest mode increasing considerably on cooling. This could be the first experimental evidence that this lowest-frequency phonon mode exhibits strong spin-phonon coupling or may couple to a continuum of magnetic excitations, inconsistent with conventional magnons. We find only marginal shifts in the eigenfrequencies when passing from the high-temperature monoclinic to the low-temperature rhombohedral phase, signaling that the molecular Cl-Ru-Cl sandwich determines the phonon properties alone, while the stacking sequence and the overall crystal symmetry seems to play a minor role. We also arrive at a value of the electronic dielectric constant $\varepsilon_\infty = 6.22$ at room temperature, which slightly increases on decreasing temperature. It is clear that these minor temperature dependent changes of the dielectric constant cannot explain the enormous changes of the reflectivity as documented in Fig. 2. Because we observed no anomalies or splittings of modes at the structural phase transition, we argue that the results listed in Table 1 represent phonon modes and complex permittivity of a single Cl-Ru-Cl layer. For the overall reflectance of $\alpha$-RuCl$_3$ the molecular stacking sequence seems to play the dominant role. However, we have to admit that the D$_{3d}$ symmetry of the stacks allows 3 in-plane modes only. On the other hand, we are not aware of any report of eigenfrequencies in the tri-chloride structure above 400 cm$^{-1}$, which raises some doubts if the high-frequency excitation is a pure one-phonon mode or is of different origin.

TABLE 1: Phonon properties of $\alpha$-RuCl$_3$ with the incident light perpendicular to the molecular stacks, at room temperature and at 10 K. Eigenfrequencies $\omega_0$, oscillator strengths $\Delta\varepsilon$ and damping coefficients $\gamma$ are listed. Also indicated is the dipolar strength of all IR active phonon modes and the high-frequency electronic dielectric constant $\varepsilon_\infty$.

| 295 K $\varepsilon_\infty = 6.22$ | | | 10 K $\varepsilon_\infty = 6.33$ | | |
|---|---|---|---|---|---|
| $\omega_0$ (cm$^{-1}$) | $\Delta\varepsilon$ | $\gamma$ (cm$^{-1}$) | $\omega_0$ (cm$^{-1}$) | $\Delta\varepsilon$ | $\gamma$ (cm$^{-1}$) |
| 121.5 | 0.01 | 37.9 | 122.8 | 0.01 | 53.9 |
| 167.1 | 0.02 | 25.6 | 171.3 | 0.006 | 9.4 |
| 311.5 | 1.21 | 4.5 | 306.9 | 0.97 | 2.2 |
| 575.7 | 0.01 | 8.6 | 586.8 | 0.18 | 0.4 |
| | $\Sigma \Delta\varepsilon = 1.25$ | | | $\Sigma \Delta\varepsilon = 1.17$ | |

In the first IR work [8], published 50 years ago, eigenfrequencies close to 188 and 315 cm$^{-1}$ were reported and were interpreted as stretching and deformation of Ru-Cl bonds, respectively. A weak absorption close to 274 cm$^{-1}$ was not further discussed [8], and does not correspond to any of our observed phonon modes. In a recent work by Little et al. [39], two modes with eigenfrequencies close to 167 and 284 cm$^{-1}$ were identified, when analyzing room-temperature transmission data. In recent Raman scattering experiments [12,14], a series of phonons were detected, which are located between 100 and 400 cm$^{-1}$. In the detailed study of Ref. [14] the authors identified a total of eight Raman modes. The temperature dependence of the Raman shifts of these modes showed changes in the order of 1% in passing the structural phase transition. Similar effects would be unobservable and well within experimental uncertainties in the present IR experiments.

## C. Electronic transitions

We now turn to wave numbers beyond the phonon bands, to investigate the low-energy electronic DOS. The main problem to model and understand low-lying (< 1 eV) electronic transitions in α-RuCl$_3$ certainly is the fact that it is unclear, if a localized electron picture, dominated by on-site, atomic-type transitions, or a band picture characteristic for intersite electronic transitions, is appropriate for the low-energy electron DOS. One should be aware that the temperature dependent resistivity of α-RuCl$_3$ was described by an activated behavior with a band gap of order 100 – 300 meV [32,34]. The character of the DC resistivity provides a first important hint on the size and the character of the band gap. We reanalyzed the published DC resistivity results [32,34,39] and conclude that the temperature dependent resistivity, for temperatures 100 K < T < 300 K can well be described by a purely thermally activated behavior, ρ ~ exp(E$_g$/2 k$_B$T), with an energy barrier of approximately 200 meV. The electrical resistance below room temperature, as documented in Ref. [32] and Ref. [34] reveals a canonical semiconducting behavior, with no indications of hopping transport of localized charge carriers. It is unclear if any type of excitonic or on-site electronic excitations can be responsible for this type of canonical semiconducting resistance and we argue that a band picture with electronic intersite interactions is the better starting point for the low-energy electronic DOS.

Figure 5 provides a sketch of the limiting electronic structures, assuming pure either on-site or inter-site (band-like) electronic transitions. The starting point for α-RuCl$_3$ is the ruthenium d$^5$ electronic configuration in octahedral symmetry, which exhibits a CF splitting into a lower t$_{2g}$ and an upper e$_g$ level with a characteristic splitting of 10 Dq ~ 2 eV. The large CF splitting prompts a low-spin configuration of the ruthenium spins. The ground state is split by spin-orbit coupling into a J = 3/2 ground quartet and a J = ½ excited doublet, with a coupling constant λ ~ 100 meV [Fig. 5(a)]. To be more precise, the spin-orbital splitting of the ground state was determined as 3λ/2 ~ 195 meV by neutron scattering [15] and as 3λ/2 ~ 145 meV by Raman scattering [35]. The lowest electronic excitations in an atomic-like on-site picture are expected into the SOC split intermediate-spin state t$_{2g}^4$e$_g^1$ [Fig. 5(b)]. These excitations are expected at 300, 500, and 700 meV [35] and are compatible with early observations by Binotto et al. [32] and Guizetti et al. [33]. Detailed calculations of the t$_{2g}^m$e$_g^n$ have been published by Yadav et al. [$^{50}$

A number of band-structure calculations are published in recent literature [30,37,38,51] and all agree that incorporating SOC still results in a metallic ground state of α-RuCl$_3$, with the Fermi energy at the upper edge of the t$_{2g}$ band. Electronic correlations are necessary to provide an insulating ground state. Here the J = ½ states of the spin-orbit coupled t$_{2g}$ electrons form the lower (LHB) and upper

Hubbard band (UHB), separated by a band gap of the order of 200 meV, while the remaining J = 3/2 states yield the bulk of the $t_{2g}$ electrons, 1 eV below the Fermi energy. The schematic electronic DOS, as shown in Fig. 5 (c), is analogous to results derived from band structure calculations from Plumb et al. [37], Johnson et al. [30], and Kim et al. [38]. However, it has to be mentioned that LDA+ SOC + U calculations with a different set of parameters, including a larger Hubbard U, result in a band gap of 1.9 eV, coinciding with results from photoemission and inverse photoemission spectroscopy [51]. Support for a small Hubbard band gap comes from the temperature dependence of the electrical resistance [32,34] and from tunneling techniques that observe an energy gap of 250 meV [52]. If we assume an intrinsic energy gap between valence and conduction band of order 200 meV, which is consistent with temperature dependent measurements of the electrical resistance, then we think that the Mott-Hubbard type band structure as shown in Fig. 5 (c) is a good starting point for the low-lying electronic DOS in α-RuCl₃.

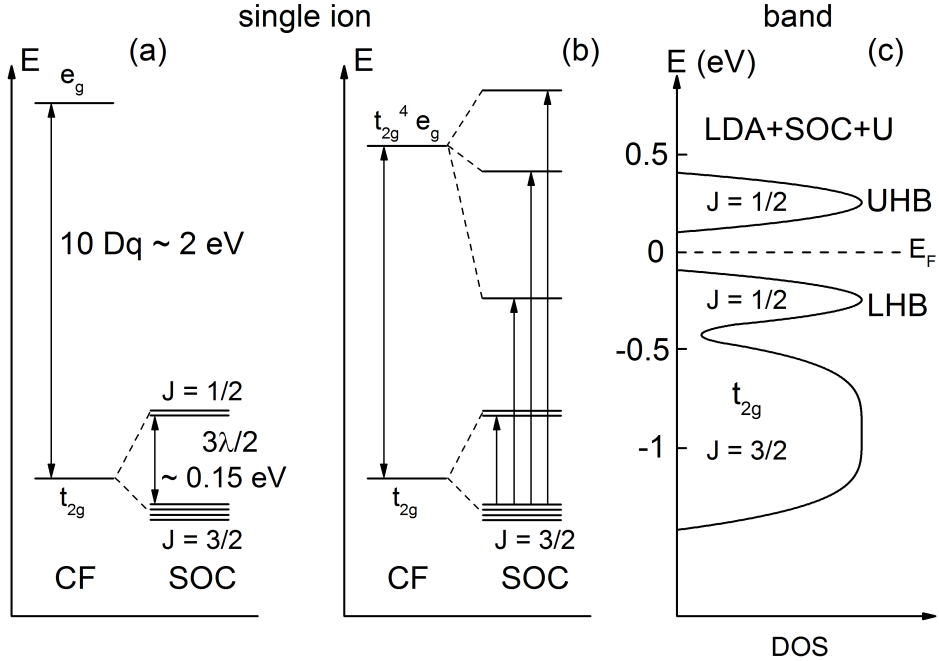

FIG.5: Schematic electronic structure of the ruthenium $d^5$ electrons in an octahedral crystal field. (a): Crystal field splitting and spin-orbit coupling. (b): The lowest electronic transitions from the spin-orbit split ground state into $t_{2g}^4 e_g^1$ levels. (c): Schematic electronic density of states in the band picture calculated including spin-orbit coupling and electronic correlations with a Hubbard U. The J = ½ states of the $t_{2g}$ electrons are split into a lower (LHB) and upper Hubbard band (UHB). This schematic band structure is reproduced from published band-structure calculations as described in the text. For the band structure, an approximate energy scale is indicated. A similar schematic electronic density of states can be found in Ref. [38].

We wanted to further elucidate this problem, by exploring the onset of the absorptivity just beyond the phonon frequencies. At low temperatures, the transmittance beyond 1200 cm⁻¹, corresponding to 150 meV, abruptly goes to zero. We interpret this step like dramatic decrease of the transmittance as onset of electronic excitations from the lower to the upper Hubbard band. If we follow the temperature dependence of the transmission beyond the phonon bands, as indicated in Fig. 3, we observe a strong temperature dependence of the onset of the band gap. Assuming that the transition from the lower to the upper Hubbard band is a direct transition, the absorption coefficient α is described by $\alpha\omega \sim (\hbar\omega - E_g)^n$, with n = 1/2 for a dipole allowed, and n = 3/2 for a dipole forbidden transition [53]. As a result, forbidden absorption edges are more difficult to identify. We were not able to detect any band-gap features in the reflectivity (see inset of Fig. 1), hence, we assume the band gap between LHB and UHB to be dipole forbidden.

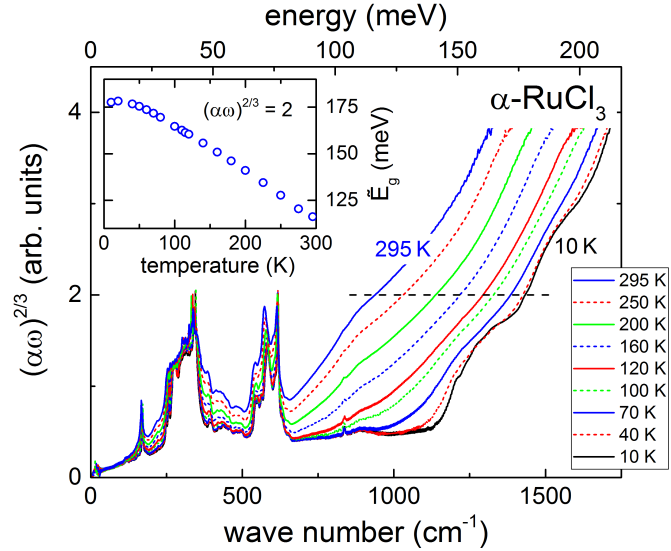

FIG. 6: Wave number dependence of absorption in α-RuCl$_3$ with the incident light perpendicular to the molecular stacks, focusing on the onset of the correlation-induced band gap at selected temperatures between 10 and 295 K. The experimental results are plotted to linearize the absorptivity for a direct forbidden electronic transition. To indicate the temperature dependence of the apparent band gap $\tilde{E}$, the inset shows a cut at constant $(\alpha\omega)^{2/3}$, indicated by the horizontal dashed line in the main frame.

The character of an electronic transition can be unambiguously determined by studying the onset of the absorption coefficient in more detail. Our results in α-RuCl$_3$ are documented in Fig. 6. Here we plotted the absorptivity vs. wave number to linearize the experimental results. For all temperatures, this linearization assuming an exponent of n = 3/2 works rather well, providing some arguments in favor of a direct forbidden dipole transition. The wavy structure of the absorption coefficient close to the band edge indicates the importance of additional phonon processes in this electronic transition, which always are relevant in direct forbidden dipole transitions. At 10 K, still in the low-temperature paramagnetic regime, we find a clear-cut onset of the correlation-induced band gap close to 150 meV. On increasing temperatures, the band gap closes and at room temperature, the band gap evolves directly beyond the phonon bands. To avoid any ambiguities introduced by a possible erroneous extrapolation, we determined the temperature dependence of the band gap at constant absorptivity. This so determined band gap, which we call apparent gap $\tilde{E}$, will always be slightly larger than the true band gap $E_g$, determined via an extrapolation to zero absorptivity. In the inset of Fig. 6 we follow the temperature evolution of $\tilde{E}$ by measuring the absorptivity at a constant value, as indicated by the horizontal dashed line in Fig. 6. Indeed, this apparent band gap increases by a factor of two between 10 K and room temperature. It is yet unclear, whether this strong temperature dependence of the band gap, which corresponds to a strong blue shift of the optical gap on decreasing temperatures, can be explained in terms of a correlation-induced gap or whether this experimentally observed temperature dependence rather indicates on-site single ion interactions in a temperature dependent crystal electric field. It is also clear that the structural phase transition cannot be identified in the temperature dependence of the gap. We conclude that the band gap in α-RuCl$_3$ is determined by the symmetry of the molecular stacks and not by the stacking sequence.

To get an overview of the complex dielectric permittivity of α-RuCl$_3$, we combined our THz data, where we measured transmission and phase shift, with our reflectivity and transmission results below 8000 cm$^{-1}$. We converted these results into spectra of the complex dielectric permittivity, dielectric constant and dielectric loss, and into the real part of the optical conductivity. The wave-number dependent dielectric constant is shown in Fig. 7 (a), the dielectric loss in Fig. 7 (b), and finally the real part of the optical conductivity in Fig. 7 (c). Here, the solid line results from experiments performed in the course of this work. The dashed line in Fig. 7 (c) represent the optical conductivity as determined from the THz results, multiplied by a factor of 50. On the scale of Fig. 7 (c) the optical conductivity would be almost invisible. In time-domain THz spectroscopy transmission and phase shift are measured simultaneously and allow determining real and imaginary parts of the dielectric function directly with high precision. The THz results agree well with the FIR results, where the absolute values have been determined from transmission experiments. As discussed earlier, the reflectivity had to be scaled by a factor of 2.58 to be compatible with these results as shown in Fig. 7.

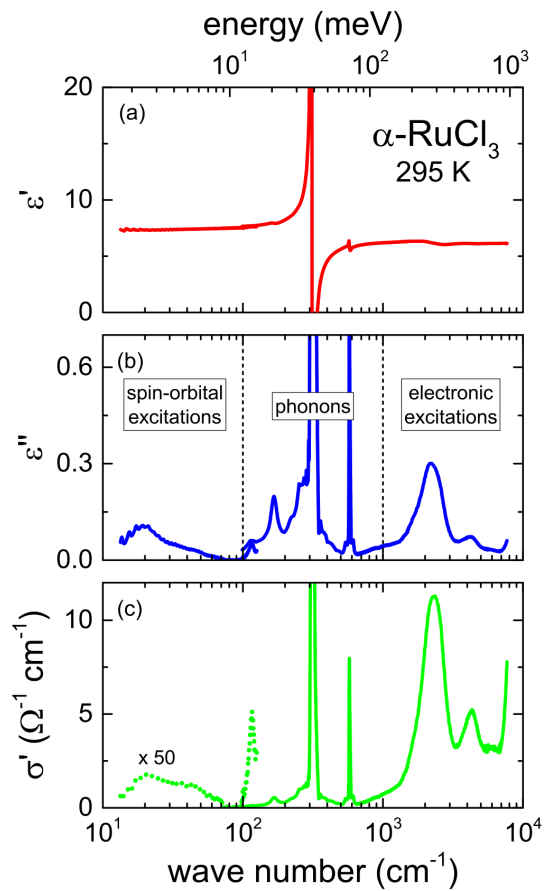

FIG. 7: Wave number dependence of (a) dielectric constant, (b) dielectric loss and (c) real part of the optical conductivity in α-RuCl$_3$ at room temperature. All spectra were measured with the incident light perpendicular to the molecular stacks. The spectra plotted with solid lines were derived from THz results (< 120 cm$^{-1}$), as well as from reflectivity and transmission experiments in the FIR and MIR range ( > 100 cm$^{-1}$) in the present experiments. The conductivity in the THz regime has been multiplied by a factor of 50. Otherwise, it would be unobservable using a linear scale. There is almost perfect match of the THz and FIR results in the dielectric constant and only a minor mismatch in the dielectric loss. In both experimental set-ups the weak phonon-like anomaly close to 120 cm$^{-1}$ can be identified. This anomaly becomes pronounced in the conductivity shown in frame (c). The dashed vertical lines in (b) indicate the three regimes corresponding to a different nature of the observed excitations: Spin-orbital excitations at THz frequencies, phonons in the FIR regime and electronic transitions at MIR frequencies.

The loss spectrum of α-RuCl$_3$ can be subdivided into three regimes: spin-orbital excitations from 10 to 100 cm$^{-1}$, phonons from 100 – 1000 cm$^{-1}$ and finally, low-energy electronic transitions from 1000 to 10.000 cm$^{-1}$. We think that the room temperature THz peak results from transitions within the J = 3/2 ground state. We assume a minor splitting of the ground-state quartet, which could result from deviations of the local crystal field from cubic symmetry. This broad hump certainly hampers any indications of a power-law behavior of the optical conductivity, as has been reported for the spin-liquid candidates Herbertsmithite [41]. At FIR frequencies, four phonon modes are visible, which have been analyzed earlier. Finally, low-lying electronic excitations immediately follow the phonon bands. The strong reststrahlen band fully dominates the dielectric constant. All the other excitations are hardly detectable and the dielectric constant is markedly featureless below, as well as above this dominating phonon mode.

There exist severe problems with the observation of 4 phonon bands at all temperatures, as documented in Fig. 7 and in Tab. 1. From symmetry considerations 4 IR active in-plane modes are expected in the monoclinic high-temperature phase, with an ABC stacking sequence, but only 3 in-plane modes should occur in the rhombohedral low-temperature phase with AB stacking. 3 in-plane modes are expected assuming a sandwich structure with independent stacks. First, no mode vanishes at the structural phase transition, a fact providing some arguments for independent molecular stacks dominating the phonon response. In this case, either we observe one mode with out-of-plane symmetry due to a slight misalignment of the sample or the highest frequency mode close to 580 cm$^{-1}$ is no phonon mode. As indicated earlier, we are not aware of any report of phonon modes in the tri-chlorides with frequencies higher than 400 cm$^{-1}$. This is true for experiments and calculations of chromium compounds [46,54] and results from first-principle calculations of the phonon density of states in non-magnetic RhCl$_3$ [55]. A way out of this controversy would be the assumption that this rather narrow excitation corresponds to the transition in the ground state split by SOC. This spin-orbit exciton in α-RuCl$_3$ has been determined by Raman scattering to have an excitation energy close to 144 meV, corresponding to 1160 cm$^{-1}$ [35] and was found close to 195 meV (~ 1570 cm$^{-1}$) in inelastic neutron-scattering experiments [15].

Now we turn to the electronic transitions. So far, the significant peaks at 2000 and 4000 cm$^{-1}$, visible in Figs. 7 (b) and (c), followed by a weak excitation close to 6000 cm$^{-1}$, were interpreted as parity and spin forbidden transitions between the Ruthenium derived d bands [35]. These excitations should be extremely low in intensity and indeed, these excitations are unobservable in the MIR reflectivity (inset Fig. 1). In specific situations, the spectral weight of such d-d transitions can be borrowed from allowed d-p transitions, which could be the case in α-RuCl$_3$. The peak at located at 2000 cm$^{-1}$ roughly corresponds to the energy of the band gap, but in the interpretation given above is a purely on-site local excitation. It is unclear if this local excitation can be responsible for bulk conductivity, which following hand-waving arguments should stem from electronic intersite transitions.

On the other hand, it is unclear if this loss spectrum [Fig. 7 (b)] or this dynamic conductivity [Fig. 7 (c)], is compatible with electronic transitions from the density of states as indicated in Fig. 5(c). One possibility would be that this fine structure results from electronic transitions from an electronic density of states, as derived from first-principle calculations by Johnson et al. [30], with a number of narrow peaks of the t$_{2g}$ electronic density of states just below the Fermi energy. Furthermore, the role of on-site d-d excitations within the band gap has yet to be clarified in strongly correlated metals. To arrive at final conclusions, one has to await future realistic model calculations of dielectric constant and loss.

## IV. CONCLUSIONS

In the present work, we performed time domain THz spectroscopy as well as transmittance and reflectance experiments in the FIR and MIR frequency regime of the Kitaev candidate material $\alpha$-RuCl$_3$ for temperatures from 5 to 295 K. We present broadband spectra of dielectric constant and dielectric loss. In accordance with earlier work [14,26,27], we identified a first-order structural phase transition with an unusual large hysteresis extending from 60 to 166 K. This phase transition is accompanied by dramatic changes of the reflectivity, similar to observations in phase-change materials [45]. We speculate that this tremendous temperature-induced reflectivity variations result from details of the layering of the molecular Cl-Ru-Cl stacks, changing from an ABC sequence in the monoclinic phase at room temperature, to an AB stacking sequence in the low-temperature rhombohedral phase. We observed four phonon modes at room temperature. Eigenfrequencies and intensities of all phonon excitations reveal only minor changes as function of temperature and remain almost unchanged when passing the first-order phase transition. No mode splitting or additional modes could be identified. This fact documents that these phonon modes originate from the molecular stacks and can be described by the $D_{3d}$ symmetry of these molecular units only. The lowest mode at 121.5 cm$^{-1}$ is heavily damped at room temperature and this damping (with large experimental uncertainties) strongly increases on decreasing temperature. This behavior could signal the coupling of this phonon mode to a broad continuum of unconventional magnetic excitations as observed by Raman spectroscopy in [14]. Further work is necessary to elucidate this strange behavior. However, in $D_{3d}$ symmetry only 3 in-plane modes are expected and the observation of an excitation at 580 cm$^{-1}$ remains a mystery. No such high-frequency phonon mode was observed in any tri-chloride system studied so far, hence, it could be of electronic origin. If one assumes a transition between the spin-orbit split ground state, the excitation energy seems to be too small compared to published results [15,35] and would result in a rather small spin-orbit coupling constant $\lambda \sim 50$ meV.

At THz frequencies, we identified a broad structure close to 20 cm$^{-1}$, which could result from transitions between the ground-state levels of the d$^5$ electronic configuration in the crystals field. The ground state quartet could be split by local deviations of the crystal field from cubic symmetry. This low-frequency intensity is only weakly temperature dependent and has to be taken into consideration when additional fractionalized excitations should be identified. We also studied the temperature evolution of the low-energy electronic density of states. The frequency dependent onset of the absorption coefficient can well be described assuming a direct forbidden electric dipole transition. We speculate that this transition describes an electronic transition from the lower into the upper Hubbard band, built up by a fraction of $t_{2g}$ electrons mainly with J = ½ character. The band gap perfectly matches results obtained from electrical DC resistivity experiments. This electronic transition strongly depends on temperature and shows a strong blue shift on decreasing temperature. However, the fine structure of the electron density of states can also be explained in terms of parity and spin forbidden d-d transitions. It is unclear if these local excitations can explain bulk conductivity with a small gap.

## ACKNOWLEDGEMENTS

We thank Roser Valenti and Igor Mazin for stimulating discussions. We thank P. Lunkenheimer for help with the design of the figures. This research was partly funded by Deutsche Forschungsgemeinschaft DFG via the Transregional Collaborative Research Center TRR 80 "From Electronic correlations to functionality" (Augsburg, Munich, Stuttgart) and by a Korea Research Foundation (KRF) Grant, funded by the Korean Government (MEST) (Grant No.: 20160874).

[1] L. Balents, Nature **464**, 199 (2010).

[2] A. Kitaev, Ann. Phys. **321**, 2 (2006).

[3] G. Jackeli and G. Khalliullin, Phys. Rev. Lett. **102**, 017205 (2009).

[4] S. M. Winter, Y. Li, H. O. Jeschke, and R. Valenti, Phys. Rev. B **93**, 214431 (2016).

[5] Y. Singh and P. Gegenwart, Phys. Rev. B **82**, 064412 (2010).

[6] Y. Singh, S. Manni, J. Reuther, T. Berljin, R. Thomale, W. Ku, S. Trebst, and P. Gegenwart, Phys. Rev. Lett. **108**, 127203 (2012).

[7] J. M. Fletcher, W. E. Gardner, E. W. Hooper, K. R. Hyde, F. H. Moore, and J. L. Woodhead, Nature **199**, 1089 (1963).

[8] J. M. Fletcher, W. E. Gardner, A. C. Fox, and G. Topping, J. Chem Soc. (A) 1038 (1967).

[9] G. Baskaran, S. Mandal, and R. Shankar, Phys. Rev. Lett. **98**, 247201 (2007).

[10] J. Knolle, D. L. Kovrizhin, J. T. Chalker, and R. Moessner, Phys. Rev. Lett. **112**, 207203 (2014).

[11] J. Knolle, G.-W. Chern, D. L. Kovrizhin, R. Moessner, and N. B. Perkins, Phys. Rev. Lett. **113**, 187201 (2014).

[12] L. J. Sandilands, Y. Tian, Kemp W. Plumb, Y.-J. Kim, and K. S. Burch, Phys. Rev. Lett. **114**, 147201 (2015).

[13] J. Nasu, J. Knolle, D. L. Kovrizhin, Y. Motome, and R. Moessner, Nature Physics **12**, 912 (2016).

[14] A. Glamazda, P. Lemmens, S.-H. Do, Y. S. Kwon, and K.-Y. Choi, Phys. Rev. B **95**, 174429 (2017).

[15] A. Banerjee, C, A, Bridges, J.-Q. Yan, A. A. Aczel, L. Li, M. B. Stone, G. E. Granroth, M. D. Lumsden, Y. Yiu, J. Knolle, S. Bhattarcherjee, D. L. Kovrizhin, R. Moessner, D. A. Tennant, D. G. Mandrus, and  S. E. Nagler,  Nat. Materials **15**, 733 (2016).

[16] K. Ran, J. Wang, W. Wang, Z.-Y.Dong, X. Ren, S. Bao, S. Li, ZU. Ma, Y. Gan, Y. Zhang, J. T. Park, G. Deng, S. Danilkin, S.-L. Yu, J.-X. Liand, and J. Wen, Phys. Rev. Lett. **118**, 107203 (2017).

[17] S.-H. Do, S.-Y. Park, J. Yoshitake, J. Nasu, Y. Motome, Y. S. Kwon, D. T. Adroja, D. J. Voneshen, K. Kim, T.-H. Jang, J.-H. Park, K.-Y. Choi, and S. Ji, unpublished, arXiv: 1703.0108 (2017).

[18] S. M. Winter, K. Riedel, A. Honecker, and R. Valenti, unpublished, arXiv: 1702.08466 (2017).

[19] E. V. Stroganov and K. V. Ovchinnikov, Vestnik. Leningrad. Univ. Ser. Fiz. I Khim **12**, 22 (1957).

[20] K. Brodersen, F. Moers, and H. G. Schnering, Naturwissenschaften **52**, 205 (1965).

[21] K. Brodersen, G. Thiele, H. Ohnsorge, I. Recke, and F. Moers, J. Less-Common Met. **15**, 347 (1968).

[22] H.-J. Cantow, H. Hillebrecht, S. N. Magonov, H. W. Rotter, M. Drechsler, and G. Thiele, Angew. Chem. Int. Ed. **29**, 537 (1990).

[23] B. Morosin and A. Narath, J. Chem. Phys. **40**, 1958 (1964).

[24] M. A. McGuire, H. Dixit, V. R. Cooper, and B. C. Sales, Chem. Mater. **27**, 612 (2015).

[25] M. A. McGuire, G. Clark, S. KC, W. M. Chance, G. E. Jellison Jr., V. R. Cooper, X. Xu, and B. C. Sales, unpublished, arXiv: 1706.01796 (2017).

[26] Y. Kubota, H. Tanaka, T. Ono, Y. Narumi, and K. Kindo, Phys. Rev. B **91**, 094422 (2015).

[27] S. Y. Park, S.-H. Do, K.-Y. Choi, D. Jang, T.-H. Jang, J. Schefer, C.-M. Wu, J. S. Gardner, J. M. S. Park, J.-H. Park, and S. Ji, unpublished, arXiv: 1609.05690 (2016).

[28] Y. Kobayashi, T. Okada, K. Asai, M. Katada, H. Sano, and F. Ambe, Inorg. Chem. **31**, 4570 (1992).

[29] J. A. Sears, M. Songvilay, K. W. Plumb, J. P. Clancy, Y. Qiu, Y. Zhao, D. Parshall, and Y.-J. Kim, Phys. Rev. B **91**, 144420 (2015).

[30] R. D. Johnson, S. C. Williams, A. A. Haghighirad, J. Singleton, V. Zapf, P. Manuel, I. I. Mazin, Y. Li, H. O. Jeschke, R. Valentí, and R. Coldea, Phys. Rev. B **92**, 235119 (2015).

[31] H. B. Cao, A. Banerjee, J.-Q. Yan, C. A. Bridges, M. D. Lumsden, D. G. Mandrus, D. A. Tennant, B. C. Chakoumakos, and S. E. Nagler, Phys. Rev. B **93**, 134423 (2016).

[32] L. Binotto, I. Pollini, and G. Spinolo, phys. stat. sol. (b) **44**, 245 (1971).

[33] G. Guizzetti, E. Reguzzoni, and I. Pollini, Phys. Lett. **70A**, 34 (1979).

[34] S. Rojas and G. Spinolo, Sol. Stat. Commun. **48**, 349 (1983).

[35] L. J. Sandiland, Y. Tian, A. A. Reijnders, H.-S. Kim, K. W. Plumb, Y.-J. Kim, H.-Y. Kee, and K. S. Burch, Phys. Rev. B **93**, 075144 (2016).

[36] L. J. Sandilands, C. H. Sohn, H. J. Park, S. Y. Kim, K. W. Kim. J. A. Sears, Y.-J. Kim, and T. W. Noh, Phys. Rev. B **94**, 195156 (2016).

[37] K. W. Plumb, J. P. Clancy, L. J. Sandilands, V. Shankar, Y. F. Hu, K. S. Burch, H.-Y. Kee, and Y.-J. Kim, Phys. Rev. B **90**, 041112 (R) (2014).

[38] H.-S. Kim, V. Shankar, A. Catuneanu, and H.-Y. Kee, Phys. Rev. B **91**, 241110 (R) (2015).

[39] A. Little, Liang Wu, P. Lampen-Kelley, A. Banerjee, S. Pantankar, D. Rees, C. A. Bridges, J.-Q. Yan, D. Mandrus, S. E. Nagler, and J. Orenstein, unpublished, arXiv: 1704.07357 (2017).

[40] Zhe Wang, S. Reschke, D. Hüvonen, S.-H. Do, K.-Y. Choi, M. Gensch, U. Nagel, T. Room, and A. Loidl, unpublished, arXiv: 1706.06157 (2017).

[41] D. V. Pilon, C. H. Lui, T.-H. Han, D. Shrekenhamer, A. J. Frenzel, W. J. Padilla, Y. S. Lee, and N. Gedik, Phys. Rev. Lett. **111**, 127401 (2013).

[42] A. B. Kuzmenko, Rev. Sci. Instr. **76**, 083108 (2005).

[43] A. B. Kuzmenko, "RefFIT v. 1.2.99," University of Geneva (2016), https://sites.google.com/site/reffitprogram/home.

[44] J. M. Skelton, K. Kobayashi, Y. Sutou, and S. R. Elliott, Appl. Phys. Lett. **102**, 224105 (2013).

[45] K. Shportko, S. Kremers, M. Woda, D. Lencer, J. Robertson, and M. Wuttig, Nat. Materials **7**, 653 (2008).

[46] V. M. Bermudez, Sol. Stat. Commun. **19**, 693 (1976).

[47] J. A. Harrington and M. Haas, Phys. Rev. Lett. **31**, 710 (1973).

[48] J. T. Gourley and W. A. Runciman, J. Phys. C: Sol. Stat. Phys. **6**, 583 (1973).

[49] B. Bendow, H. G. Lipson, S. S. Mitra, Phys. Rev. B **20**, 1747 (1979).

[50] R. Yadav, N. A. Bogdanov, V. M. Katukuri, S. Nishimoto, J. van den Brink, and L. Hozoi, Sci. Reports **6**, 37925 ((2016).

[51] S. Sinn, C. H. Kim, B. H. Kim, K. D. Lee, C. J. Won, J. S. Oh, M. Han, Y. J. Chang, N. Hur, H. Sato, B.-G. Park, C. Kim, H.-D. Kim, and T. W. Noh1, Sci. Rep. **6**, 39544 (2016).

[52] M. Ziatdinov, A. Banerjee, A. Maksov, T. Berlijn, W. Zhou, H. B. Cao, J.-Q. Yan, C. A. Bridges, D. G. Mandrus, S. E. Nagler, A. P. Baddorf, and S. V. Kalinin, Nature Comm. **7**, 13774 (2016).

[53] N. F. Mott and E. A. Davis, Electronic Processes in Non-Crystalline Materials, 2$^{nd}$ edn. (Clarendon Press, Oxford, 1979).

[54] I. Kanesaka, H. Kawahara, A. Yamzaki, and K. Kawai, J. Mol. Structure **146**, 41 (1986).

[55] A. U. B. Wolter, L. T. Corredor, L. Jannssen, K. Nenkov, S. Schönecker, S.-H Do, K.-Y. Choi, R. Albrecht, J. Hunger, T. Doert, M. Vojta, and B. Büchner, Phys. Rev. **96**, 041405(R) (2017).